\documentclass[graybox]{svmult}

\usepackage{mathptmx}       
\usepackage{helvet}         
\usepackage{courier}        
\usepackage{type1cm}        
\usepackage{makeidx}         
\usepackage{graphicx}        
\usepackage{multicol}        
\usepackage[bottom]{footmisc}

\makeindex             

\begin{document}

\title*{Simulations of shell galaxies with GADGET-2: Multi-generation shell systems}
\titlerunning{Multi-generation shell systems}
\author{Kate\v{r}ina Barto\v{s}kov\'{a}, Bruno Jungwiert, Ivana Ebrov\'{a}, Lucie J\'{i}lkov\'{a}, Miroslav K\v{r}\'{i}\v{z}ek}
\authorrunning{Barto\v{s}kov\'{a} et al.}
\institute{
Kate\v{r}ina Barto\v{s}kov\'{a} 
\at Faculty of Science, Masaryk University, Brno, Czech Republic; Astronomical Institute, Academy of Sciences of the Czech Republic, \email{katka.bartoskova@gmail.com}
\and Bruno Jungwiert, Ivana Ebrov\'{a} and Miroslav K\v{r}\'{i}\v{z}ek 
\at Astronomical Institute, Academy of Sciences of the Czech Republic; Faculty of Mathematics and Physics, Charles University in Prague
\and Lucie J\'{i}lkov\'{a} 
\at ESO Santiago, Chile; Faculty of Science, Masaryk University, Brno, Czech Republic
}

\maketitle

\vskip-1.2truein

\abstract{As the missing complement to existing studies of shell galaxies, we carried out a set of self-consistent N-body simulations of a minor merger forming a stellar shell system within a giant elliptical galaxy. We discuss the effect of a phenomenon possibly associated with the galaxy merger simulations\,---\,a presence of multiple generations of shells.}
\section*{Two-generation shell structure}
\label{sec1:multigen}
Galaxies with stellar shells are thought to be by-products of galaxy mergers \cite{Q01}. Most, though not all, previous models, e.g.~\cite{Q01,HQ02,E01,J01},  relied on test-particle simu\-la\-tions. No systematic explorations of galactic models with these shell structures as merger debris via fully self-consistent N-body simulations, naturally invol\-ving the dynamical friction and the progressive decay of the accreted galaxy, were conducted. To bridge this gap, we decided to carry out a set of self-consistent simulations of a minor merger between a giant elliptical galaxy (gE with the mass of dark matter halo 8$\cdot$10$^{\mathsf{12}}$\,M$_{\odot}$ and stellar component 2$\cdot$10$^{\mathsf{11}}$\,M$_{\odot}$) and a satellite dwarf elliptical galaxy (2$\cdot$10$^{\mathsf{10}}$\,M$_{\odot}$ in total), using the GADGET-2 code \cite{S01}. 

In order to study differences in the resulting shell system formed in differently centrally concentrated mass distributions, we prepared simulations with the gE galaxy in two versions: a~two-component Plummer and a~two-component Hernquist model, with the same effective radius. The dwarf galaxy is then released on a radial orbit with initial velocity $\sim$100\,km$/$s  and distance of 200\,kpc from the giant galaxy.

\begin{figure}[t]
\begin{center}
\includegraphics[scale=.30]{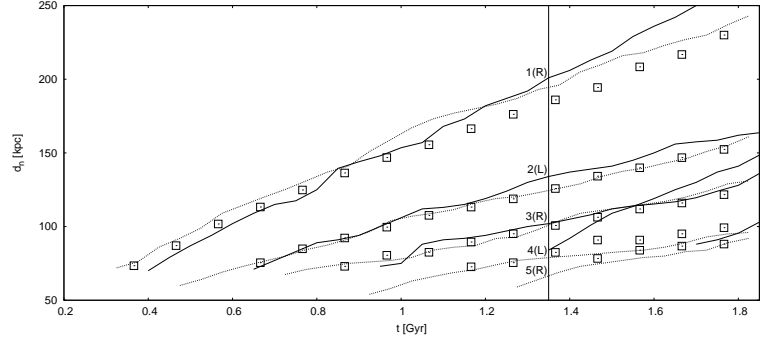}
\end{center}
\caption{The time evolution of the edges of three similar shell systems: results from the self-consistent simulation with gE galaxy modeled as the Plummer model (solid lines); from the test-particle simulation with the same initial conditions, but without the second passage of the satellite (boxes); and from the self-consistent simulation with gE galaxy represented by the Hernquist model (dotted lines). While the stars, gradually released from the potential of the satellite, oscillate in the potential of the gE galaxy, they are forming shells alternately interleaved on both sides of the gE galaxy along the merger axis\,---\,the first shell on the right side (R) is later followed by the second shell on the left (L), etc. The velocity of the shell-edge expansion\,---\,slope of $d_{{n}}(t)$ dependency\,---\,is given by the gravitational potential and by the shell ordinal number within a given generation. Therefore the first shells from each generation move with the same velocity. Such a shell galaxy arisen in the first self-consistent simulation, if observed in the particular time (e.g.~1.7~Gyr after the first passage), may appear to cause "the problem of a missing shell", since the third (originally fourth) outermost shell is then detected on the same side (L) as the second one.}
\label{fig:bartoskova1}       
\end{figure}

In the first simulation, the core of the satellite passes through, returns and makes a second passage across the center of the primary galaxy ($\sim$1\,Gyr after the first passage). This event leads to creating the second generation of shells. To our know\-ledge, this process has never been simulated in any previous study of the shell ga\-la\-xies, although predictions in this sense were made, e.g. \cite{DC02}. In the first approximation, we can look at this as a new collision between the returning core part of the satellite and the gE galaxy. Within the same generation, the shells of the debris system are moving with decreasing velocity. As the subsequent passage is not present in the latter simulation (with a two-component Hernquist model for the gE galaxy), the subsequent shells created after $\sim$1\,Gyr move with different velocities compared to those belonging to the next generation in the former simulation, see Fig.~\ref{fig:bartoskova1}.

\begin{acknowledgement}
This project is supported by grants No. 205/08/H005 (Czech Science Foundation), the project SVV 261301 (Charles University in Prague), MUNI/A/0968/2009 (Masaryk University in Brno), research plan AV0Z10030501 (Academy of Sciences of the Czech Republic) and LC06014 (Center for Theoretical Astrophysics, Czech Ministry of Education).
\end{acknowledgement}
%


\end{document}